\documentclass[superscriptaddress, amsmath, aps, prl, numerical, twocolumn, nofootinbib]{revtex4-1}

\usepackage[T1]{fontenc}
\usepackage{latexsym}
\usepackage{amssymb}
\usepackage{amsfonts}
\usepackage{amsmath}
\usepackage{amsthm}
\usepackage{multirow}
\usepackage{longtable}
\usepackage{mathtools}
\usepackage{color}
\usepackage{ulem}
\usepackage{hyperref}
\usepackage{bm}

\begin{document}

\title{Convenient location of  a near-threshold proton-emitting resonance in $^{11}$B}

\author{J. Oko{\l}owicz}
\affiliation{Institute of Nuclear Physics, Polish Academy of Sciences, Radzikowskiego 152, PL-31342 Krak{\'o}w, Poland}

\author{M. P{\l}oszajczak}
\affiliation{Grand Acc\'el\'erateur National d'Ions Lourds (GANIL), CEA/DSM - CNRS/IN2P3, BP 55027, F-14076 Caen Cedex, France}

\author{W. Nazarewicz}
\affiliation{Department of Physics and Astronomy and FRIB Laboratory,
Michigan State University, East Lansing, Michigan  48824, USA}

\begin{abstract}

The presence of cluster-like narrow resonances in the vicinity of reaction/decay thresholds is  a ubiquitous phenomenon with profound consequences. We argue that the continuum coupling, present in the  open quantum system description of the atomic nucleus, 
 can profoundly impact the nature of near-threshold states. In this Letter, we discuss the structure of the recently observed  near-threshold resonance in $^{11}$B, whose very existence explains the puzzling beta-delayed proton emission of  the neutron-rich $^{11}$Be.
\end{abstract}

\maketitle

\textit{Introduction--}
There are numerous examples of narrow resonances in light nuclei  that can be found in the proximity of particle decay thresholds. For instance, as early noticed by Ikeda et al. \cite{Ikeda1968}, $\alpha$-cluster states in light nuclei are present around 
$\alpha$-particle thresholds. Arguably the most famous state of such character is the excited $0^+$ state of $^{12}$C very close to the $\alpha$-particle separation energy, which was  postulated by Hoyle to explain production of carbon in stars
\cite{Freer2014,Freer2018}. Other examples abound \cite{Fick1978,Grancey2016,Wiescher2017,Okolowicz18}. The narrow 
near-threshold resonances are very important in astrophysical settings where most reactions happen at very low energies near the threshold \cite{Wiescher2017}.  For such states, particle emission or breakup channels can successfully compete with other decay modes, such as $\gamma$ decay.

A very unusual decay, a $\beta^-$-delayed proton decay  of a {\it neutron-rich}  nucleus $^{11}$Be, predicted theoretically in \cite{Baye2011},  was studied in
Refs.~\cite{Riisager2014,Ayyad2019}. Experimentally,  the strength of this decay mode turned out to be unexpectedly high. This puzzle was explained \cite{Riisager2014} by the presence of a narrow resonance in $^{11}$B, recently found in Ref.~\cite{Ayyad2019} slightly above the proton separation energy. As estimated in Ref.~\cite{Ayyad2019}, in order to explain the observed proton decay rate, this resonance must have a sizable single-proton content. In this Letter, we argue that the proton resonance  in $^{11}$B, which happens to be  `conveniently' located near the proton threshold is not entirely unexpected; its existence  is yet another manifestation of nuclear openness.

\textit{Near-threshold collectivity of the nuclear open quantum system--}
The pervasive appearance   of cluster states in the proximity of corresponding cluster thresholds must be a general feature, fairly independent of model details. Based on studies in the shell model embedded in the continuum (SMEC) \cite{Okolowicz2003}, it has been conjectured \cite{Okolowicz2012,Okolowicz2013} that the interplay between internal configuration mixing by nuclear interactions and external configuration mixing via decay channels leads to a new kind of near-threshold collectivity. Specifically, the proximity of the branch point singularity at the particle emission threshold induces collective mixing of shell-model  eigenstates, which results in a single `aligned eigenstate' of the system  carrying many characteristics of a nearby decay channel.

\begin{figure}[htb]
\includegraphics[width=\linewidth]{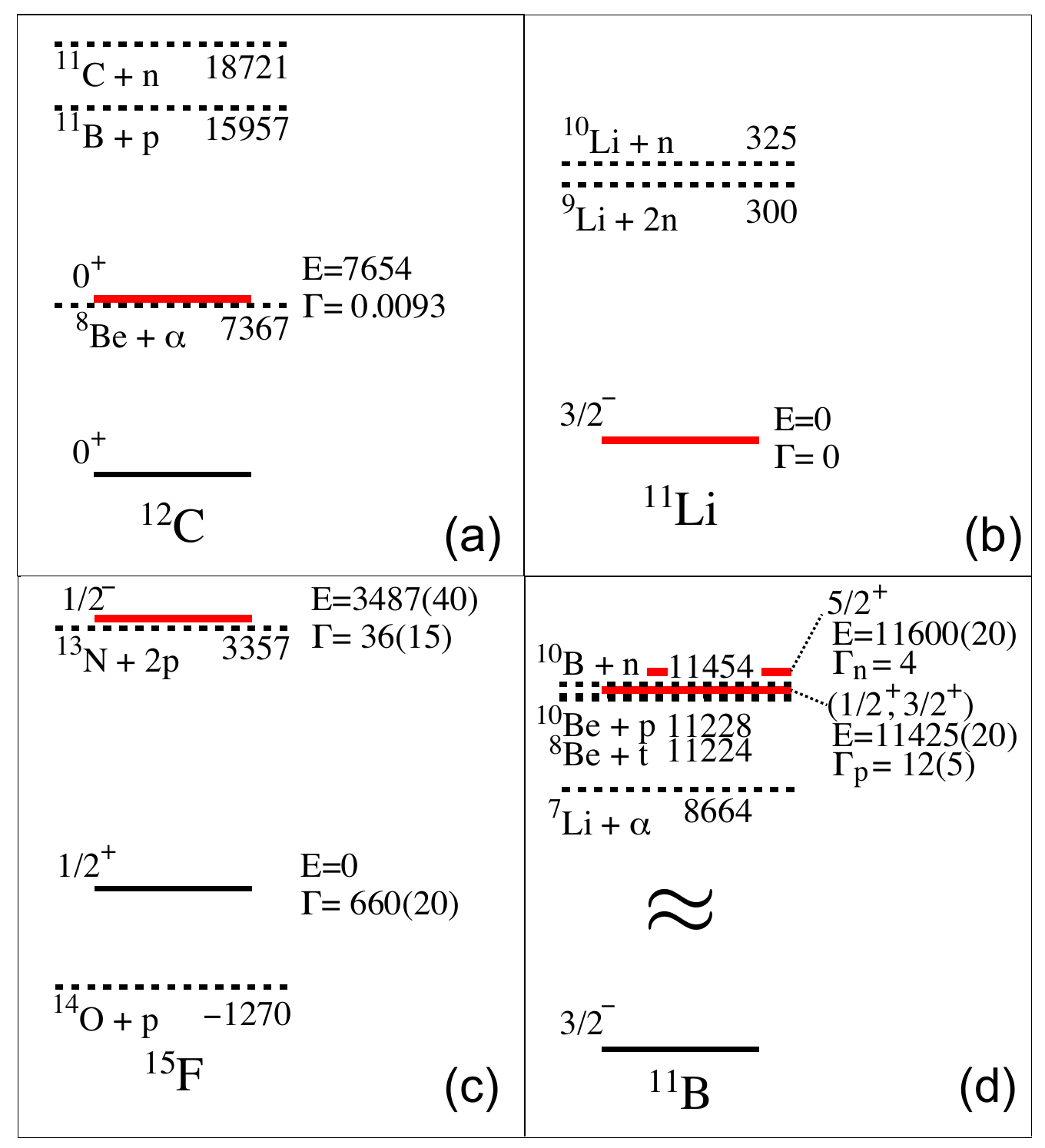}
\caption{(Color online)  Selected low-lying states and particle-decay thresholds (all in keV) in $^{12}$C \cite{NNDC} (a), $^{11}$Li \cite{NNDC} (b), $^{15}$F  \cite{Grancey2016} (c), and $^{11}$B \cite{TUNL,Ajzenberg1975,Ayyad2019} (d). The key near-threshold `aligned' states are indicated.}
\label{spectra}
\end{figure}

Figure~\ref{spectra} shows some spectacular examples of such aligned states. The Hoyle resonance  at $E=7.654$\,MeV, which is conveniently located only 287\,keV above the  $\alpha$-particle emission threshold in $^{12}$C, is believed \cite{Freer2014,Freer2018} to carry an imprint of the [$^8$Be(g.s.)$\otimes\alpha$] decay channel. The ground state (g.s.) of the Borromean halo nucleus $^{11}$Li in Fig.~\ref{spectra}(b) resembles the [$^{9}$Li(g.s.)$\otimes 2n$] configuration of the nearest $2n$-emission threshold rather  that of the [$^{10}$Li(g.s.)$\otimes n$] state \cite{Garrido1997}. (A similar situation happens in the mirror nucleus $^{11}$O, which is a $2p$ emitter \cite{Webb2019}.) Similarly, the excited $1/2^-_1$ narrow resonance in $^{15}$F, located well above the $^{14}$O+$p$ threshold and the Coulomb barrier, is expected to reflect features of a nearby 
$^{13}$N(g.s.)+$2p$ threshold;  hence, it is believed to be a $2p$ emitter  \cite{Grancey2016}. Figure~\ref{spectra}(d) shows the relevant spectrum of  $^{11}$B \cite{Ayyad2019} in the vicinity of the $^{10}$Be(g.s.)+$p$, $^{8}$Be(g.s.)+$^3$H, and 
$^{10}$B(g.s.)+$n$ thresholds \cite{TUNL}. In spite of the fact that the state at $E=11.425$\,MeV lies well above the $\alpha$ decay threshold, it does not seem to $\alpha$ decay, and it has a fairly small proton width. Indeed, its configuration is expected to resemble [$^{10}$Be$\otimes p$] rather than [$^{7}$Li$\otimes {\alpha}$]. The close proximity of proton and triton emission thresholds suggests that this resonance may also contain an admixture of [$^{8}$Be$\otimes$$^3$H] configuration.
The $5/2^+$ state at 11.6\,MeV that lies only 146\,keV above the neutron-decay threshold is also a candidate for an aligned state. Indeed, according to  Ref.~\cite{Ajzenberg1975}, this state has a vey small neutron decay width $\Gamma_n\approx 4$\,keV.

\textit{SMEC picture--}
In the simplest version of SMEC, the Hilbert space is divided into two orthogonal subspaces ${\cal Q}_{0}$ and 
${\cal Q}_{1}$ containing 0 and 1 particle in the scattering continuum, respectively. An open quantum system  description of 
${\cal Q}_0$  includes couplings to the environment of decay channels through the energy-dependent effective Hamiltonian:
\begin{equation}
{\cal H}(E)=H_{{\cal Q}_0{\cal Q}_0}+W_{{\cal Q}_0{\cal Q}_0}(E),
\label{eq21}
\end{equation}
where $H_{{\cal Q}_0{\cal Q}_0}$ denotes the standard shell-model Hamiltonian describing the internal dynamics in the closed quantum system approximation, and 
\begin{equation}
W_{{\cal Q}_0{\cal Q}_0}(E)=H_{{\cal Q}_0{\cal Q}_1}G_{{\cal Q}_1}^{(+)}(E)H_{{\cal Q}_1{\cal Q}_0},
\label{eqop4}
\end{equation}
is the energy-dependent continuum coupling term, where $E$ is a scattering energy, $G_{{\cal Q}_1}^{(+)}(E)$ is the one-nucleon Green's function, and ${H}_{{Q}_0,{Q}_1}$ and ${H}_{{Q}_1{Q}_0}$ couple ${\cal Q}_{0}$ with ${\cal Q}_{1}$. 
The effective Hamiltonian can be written as: ${\cal H}(E)=H_{{\cal Q}_0{\cal Q}_0} + V_0^2h(E)$. The energy scale in (\ref{eq21}) is defined by the lowest one-nucleon emission threshold. The channel state is defined by the coupling of one nucleon in the scattering continuum to a shell model wave function of the nucleus  $(A-1)$.

The continuum induced mixing of shell model states in a given SMEC eigenstate of ${\cal H}(E)$, $\Psi_{\alpha}$,  can be studied using the continuum-coupling correlation energy 
 \begin{equation}
E_{\rm corr}^{(\alpha)}(E)=\langle \Psi_{\alpha}| W_{{\cal Q}_0{\cal Q}_0}(E) |\Psi_{\alpha}\rangle,
\label{eq22}
\end{equation}
which can be calculated for any SMEC eigenstate.  
The point of the strongest collectivization, i.e.,the centroid of the opportunity energy window to find the aligned state, is determined by an interplay between the  Coulomb+centrifugal barrier and the continuum coupling. For higher angular momenta $\ell$ and/or for charged particle decay channels, the extremum of the continuum-coupling correlation energy is shifted above the threshold.

\textit{Results--}
In Ref.~\cite{Ayyad2019}, the  $J^{\pi}=(1/2^+,3/2^+$) resonance 
with $E=11.425(20)$\,MeV and  $\Gamma_p = 12(5)$\,keV, has been found  just 197(20)\,keV above the  one-proton emission threshold and only 29(20)\,keV below the one-neutron emission threshold. Moreover, a close-lying broad $J^{\pi}=(3/2^+$) alpha-decaying state has been suggested in Ref.~\cite{Refsgaard2019} to explain the $\beta$-delayed $\alpha$ spectrum from $^{11}$Be.

In our study of the resonance at 11.425\,MeV, the Hamiltonian $H_{{\cal Q}_0{\cal Q}_0}$ contains the  WBP$-$ interaction \cite{Yuan2017} in the full $psd$ model space. The continuum-coupling interaction was assumed to be the  Wigner-Bartlett contact force 
$V_{12}=V_0 \left[ \alpha + \beta P^{\sigma}_{12} \right] \delta\langle\bf{r}_1-\bf{r}_2\rangle$, where $\alpha + \beta = 1$ and $P^{\sigma}_{12}$ is the spin exchange operator. The spin-exchange parameter $\alpha$ has a standard value of $\alpha = 0.73$. The radial single-particle wave functions (in ${\cal Q}_0$) and the scattering wave functions 
(in ${\cal Q}_1$) are generated by the Woods-Saxon potential, which includes spin-orbit and Coulomb parts. The radius and diffuseness of the Woods-Saxon potential are $R_0=1.27 A^{1/3}$~fm and $a=0.67$~fm, respectively. The strength of the spin-orbit potential is $V_{\rm SO}=7.62$~MeV,  and the Coulomb part is calculated for a uniformly charged sphere with radius $R_0$. The depth of the central potential for neutrons and protons is adjusted to reproduce the measured separation energies of the respective $p_{3/2}$ orbits.

The calculations were carried out for $J^{\pi} = 1/2^+$ and $3/2^+$  states in $^{11}$B.  {The shell model states are mixed via the coupling to the respective one-proton}  $[{^{10}}$Be($0^+) \otimes p(s_{1/2})]^{{1/2}^{+}}$ / $[{^{10}}$Be($0^+) \otimes p(d_{3/2})]^{{3/2}^+}$, and one-neutron 
$[{^{10}}$B($3^+) \otimes n(d_{5/2})]^{J^{\pi}}$ reaction channels.

\begin{figure}[htb]
\includegraphics[width=1.0\linewidth]{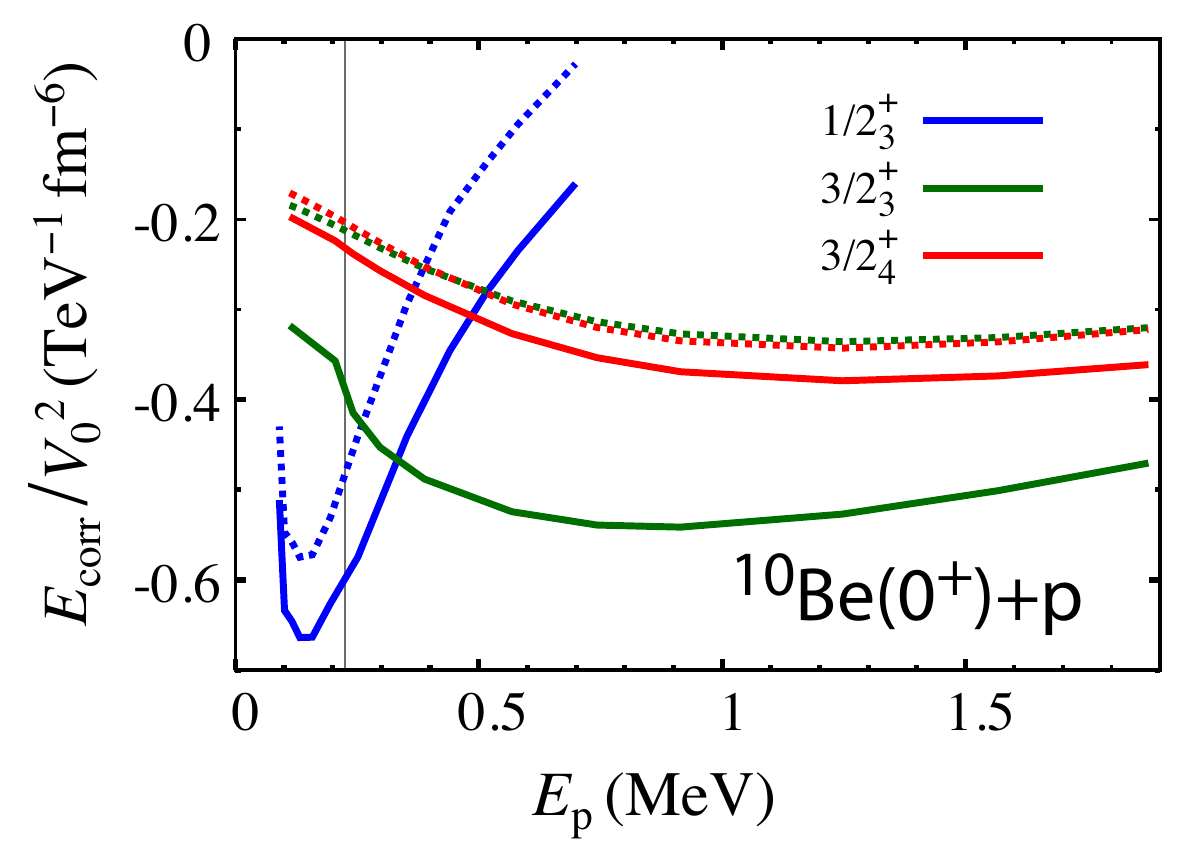}
\caption{Real part of the continuum-coupling correlation energy (\ref{eq22})
calculated in  SMEC for the  $1/2^+_3$, $3/2^+_3$, and $3/2^+_4$ states, which are in the vicinity of the experimental proton emission threshold. The results are shown  as a function of the proton energy $E_p$ in the continuum. 
Solid lines mark the calculations, which consider the coupling to both proton and neutron reaction channels. The dotted lines show the situation, in which the  coupling  to the neutron channel is  ignored.
Zero energy  corresponds to the proton decay threshold. The neutron decay threshold is marked by a thin vertical line.}
\label{ecorr}
\end{figure}

Figure~\ref{ecorr} shows the real part of the continuum-coupling correlation energy 
$E_{\rm corr}$ divided by the square of the continuum-coupling strength $V_0^2$ as a function of the proton energy $E_p$. Away from regions of avoided crossing of 
SMEC eigenstates, and in the one-channel case, $E_{\rm corr}/V_0^2(E_p)$ is a universal function of energy, independent of the continuum-coupling constant $V_0$. In order to assess the importance of one-neutron decay channels, two situations have been  considered. In the first case, denoted by the solid lines in Fig.~\ref{ecorr}, the shell model states are coupled to both one-proton  
$[{^{10}}$Be($0^+) \otimes p(s_{1/2})]^{{1/2}^+}$ / $[{^{10}}$Be($0^+) \otimes p(d_{3/2})]^{{3/2}^+}$ and one-neutron $[{^{10}}$B($3^+) \otimes n(d_{5/2})]^{J^{\pi}}$ ($J^{\pi}={1/2}^+$, ${3/2}^+$) reaction channels. In the second case shown by dotted lines, only the coupling to one-proton decay channel has been considered. At each proton energy, the contributions to the  correlation energy (\ref{eq22}) from couplings to one-proton and/or one-neutron reaction channels are calculated with the correct asymptotic of the $s_{1/2}$/$d_{3/2}$ proton and/or $d_{5/2}$ neutron single-particle state with respect to the one-proton and/or one-neutron channel threshold, respectively. This means that for each continuum energy, the depth of the Woods-Saxon potential is fixed so that the $s_{1/2}$/$d_{3/2}$ proton and/or $d_{5/2}$ neutron single-particle states are obtained at the correct energy with respect to the corresponding one-proton and/or one-neutron thresholds. 

For the $1/2^+_3$ SMEC eigenstate, all four $1/2^+$ shell-model eigenstates are coupled in the $\ell=0$ partial wave to one-proton decay channel and in the  $\ell=2$ wave to one-neutron decay channel (see Fig.~\ref{ecorr}). In this case, the strongest collectivization is predicted  at  $E_p^* \simeq 142$ keV, close to the experimental energy of the resonance \cite{Ayyad2019}. We have checked that the centroid of the opportunity window for the formation of a collective SMEC eigenstate depends weakly on the assumed charge radius of $^{11}$B. Namely, changing the radius $R_0$ by 10\% modifies $E_p^*$ by $\sim 6$\,keV. Coupling to one-neutron decay channel is weak and provides a nearly constant energy shift of the continuum-coupling correlation energy  and does not change the energy $E_p^*$.

For $3/2^+_3$ and $3/2^+_4$ SMEC eigenstates (8 states considered), one finds only very shallow minima of 
the continuum-coupling correction energy. The coupling to one-neutron decay channels 
$[{^{10}}$B($3^+) \otimes n(d_{5/2})]^{J^{\pi}}$ is very weak for $3/2^+_4$ eigenstate and the minimum of $E_{\rm corr}(E_p^*)$ in this case is seen at $E_p^* \simeq 1300$ keV. For $3/2^+_3$ eigenstate, the contributions from coupling to the one-proton and one-neutron reaction channels are of a comparable magnitude, producing a minimum of $E_{\rm corr}$ at $E_p^* \simeq 860$ keV, i.e.,  $\sim 630$ keV above the one-neutron emission threshold. In both cases, the collectivization is expected well above the experimental resonance energy; hence,  the result shown in Fig.~\ref{ecorr} strongly suggests the $J^{\pi} = 1/2^+$  assignment for the observed proton resonance.

By comparing both variants of calculations shown in Fig.~\ref{ecorr}: with and without the coupling to the neutron-decay threshold, we conclude that the coupling to the  closed one-nucleon channels does not impact our conclusions. At very low energies, the results of our SMEC calculations for the $1/2^+_3$ state are strongly affected by the presence of as many as three exceptional points in the energy interval between 88 and 92\,keV; this makes it practically impossible to identify the SMEC eigenstate below $E_p=100$\,keV.

The $5/2^+$ narrow resonance at $E=11.600(20)$\,MeV shown in Fig.~\ref{spectra}(d) that lies slightly
above the one-neutron decay threshold is known to decay by $\alpha$- and neutron-emission. The huge neutron capture cross section on $^{10}$B target at low bombarding energies is controlled by this $5/2^+$ resonance, and this suggests a large imprint of the  $[{^{10}}$B($3^+) \otimes n(d_{5/2})]^{{5/2}^+}$ reaction channel on the resonance's wave function.
In the SMEC calculation, six $5/2^+$ shell-model eigenstates are coupled in the $\ell=2$ partial wave to one-neutron decay channel. One of these states, $5/2^+_6$ , appears 
in the vicinity of one-neutron decay threshold.

\begin{figure}[htb]
\includegraphics[width=1.0\linewidth]{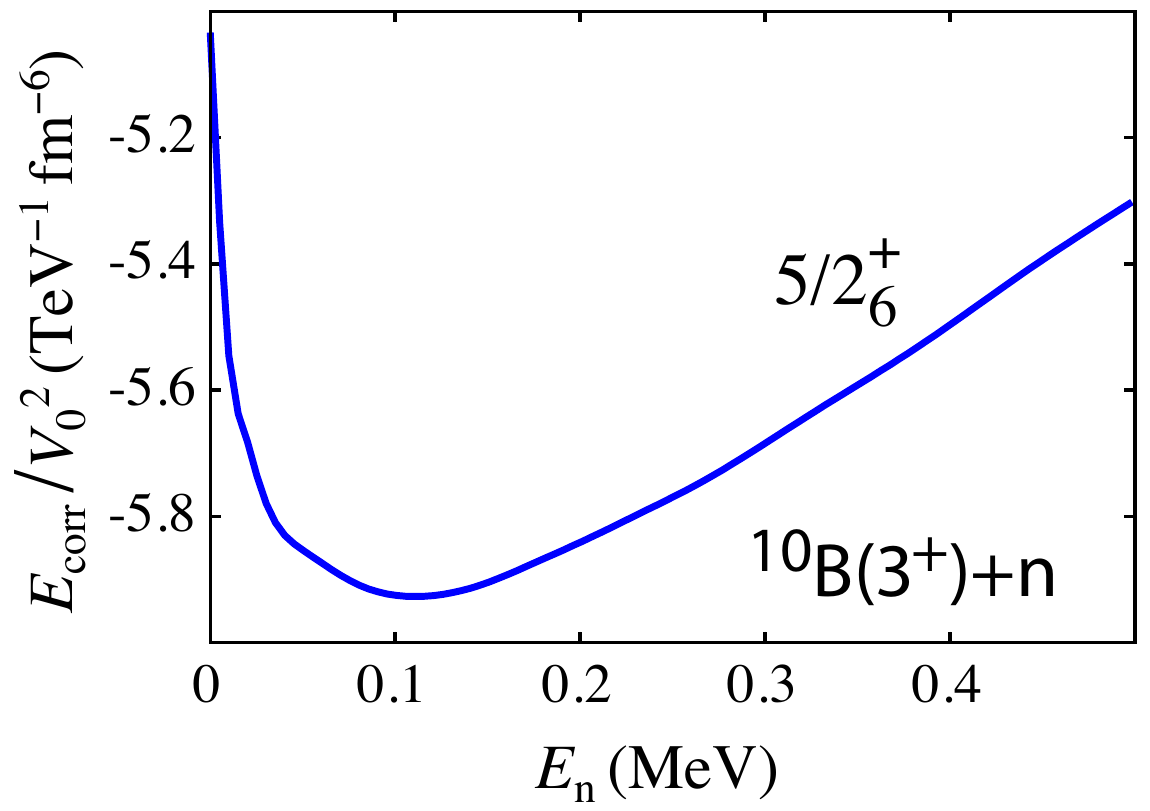}
\caption{Real part of the continuum-coupling correlation energy (\ref{eq22}) as a function of the neutron energy $E_n$ in the continuum
calculated in  SMEC for the  $5/2^+_6$ state, which lies in the vicinity of the experimental neutron decay threshold. 
Zero energy  corresponds to the neutron decay threshold. }
\label{ecorr1}
\end{figure}

Figure~\ref{ecorr1} shows the continuum-coupling correlation energy as a function of the neutron energy $E_n$ for the $5/2^+_6$ eigenstate. The coupling to the one-neutron reaction channel  $[{^{10}}$B($3^+) \otimes n(d_{5/2})]^{{5/2}^+}$ is very strong in this case. The minimum of $E_{\rm corr}(E_n^*)$ is predicted at $E_n^* = 113$\,keV, close to the experimental  energy of the $5/2^+$ resonance. However, one may notice in Fig.~\ref{ecorr1} that the continuum-coupling correlation energy is rather flat in a broad energy interval (0.09 MeV$\leq E_n \leq 0.16$ MeV) around $E_n^*$.  

\textit{Conclusions--} In this Letter, we studied the curious case of a $\beta^-p$ decay of a neutron halo nucleus $^{11}$Be through a  threshold resonance in $^{11}$B. Our SMEC calculations strongly favor the  $J^{\pi} = 1/2^+$  assignment over $3/2^+$. The wave function of the $1/2^+_3$ SMEC eigenstate carries characteristics of a nearby proton decay threshold, i.e., this state can be viewed as a core-coupled proton state [$^{10}$Be$\otimes p$] with the negligible  [$^{7}$Li$\otimes {\alpha}$] component. This conclusion is consistent with the suggestions of Refs.~\cite{Riisager2014,Yang2019} that the $\beta^-$ decay may be interpreted as a quasi-free decay of the $^{11}$Be halo neutron into a single-proton state, coupled to the $^{10}$Be core. In such scenario, the [$^{8}$Be$\otimes$$^3$H] component, if any, does not impact the $\beta^-p$ decay process.

The `alignment' of $1/2^+_3$ eigenstate with the [$^{10}$Be$\otimes p$] reaction channel also explains the large spectroscopic factor for the proton decay \cite{Ayyad2019} and very small  $\alpha$-particle decay width of this state.
The nearby  $J^{\pi}=(3/2^+$) excitation discussed in Ref.~\cite{Refsgaard2019}, on the other hand, alpha decays. A candidate for this resonance could be, e.g., the predicted 
$3/2^+_4$ state, which weakly couples to one-neutron and one-proton reaction channels.
 
Above the one-neutron  [$^{10}$B$\otimes n$] threshold, one finds a $5/2^+$ resonance, which is crucial for the neutron capture  on $^{10}$B. The neutron partial decay width for this state, $\Gamma_n=4$\,keV,  is a large if one considers the small energy above the decay threshold and the $\ell = 2$ partial wave involved in this decay. Therefore, the wave function of $5/3^+_6$ SMEC eigenstate exhibits strong collectivization by the coupling of all $5/2^+$ shell-model eigenstates  to the neutron decay threshold.

The reason for the appearance of the proton (neutron) resonances around the proton (neutron) emission threshold is the continuum coupling to the $\ell=0$ proton ($\ell=2$ neutron) scattering space. In this respect, the case of $^{11}$B follows other splendid examples of threshold states shown in Fig.~\ref{spectra}. 

Future experiments to clarify the nature of the resonance at $E=11.425$\,MeV are called for. Those include 
$^{10}$Be($p,p$)$^{10}$Be and $^{8}$Be($^{3}$H,$^{3}$H)$^{8}$Be studies.
Also, to better understand the nature of the nearby neutron reaction channel and close-lying neutron resonances, 
more experimental  and theoretical  work is needed. The former involves  $^{10}$B($d,p$)$^{11}$Be studies. The latter should clarify the impact of the virtual $\ell=0$ neutron state on the $^{10}$B+$n$ reaction channel.

\begin{acknowledgements}
Discussions with Alexandra Gade and Gregory Potel are gratefully acknowledged.
This material is based upon work supported by
the U.S. Department of Energy, Office of Science, Office of
Nuclear Physics under Award No.DE-SC0013365 (Michigan
State University) and by the COPIN and COPIGAL French-Polish scientific exchange programs.  

\end{acknowledgements}

\bibliographystyle{apsrev4-1} 
\bibliography{threshold}

\end{document}